# Scientific Workflow Applications on Amazon EC2

Gideon Juve, Ewa Deelman, Karan Vahi, Gaurang Mehta
*USC Information Sciences Institute*
{gideon,deelman,vahi,gmehta}@isi.edu

Bruce Berriman
*NASA Exoplanet Science Institute, Infrared Processing and Analysis Center, Caltech*
gbb@ipac.caltech.edu

Benjamin P. Berman
*USC Epigenome Center*
bberman@usc.edu

Phil Maechling
*Southern California Earthquake Center*
maechlin@usc.edu

## Abstract

*The proliferation of commercial cloud computing providers has generated significant interest in the scientific computing community. Much recent research has attempted to determine the benefits and drawbacks of cloud computing for scientific applications. Although clouds have many attractive features, such as virtualization, on-demand provisioning, and "pay as you go" usage-based pricing, it is not clear whether they are able to deliver the performance required for scientific applications at a reasonable price. In this paper we examine the performance and cost of clouds from the perspective of scientific workflow applications. We use three characteristic workflows to compare the performance of a commercial cloud with that of a typical HPC system, and we analyze the various costs associated with running those workflows in the cloud. We find that the performance of clouds is not unreasonable given the hardware resources provided, and that performance comparable to HPC systems can be achieved given similar resources. We also find that the cost of running workflows on a commercial cloud can be reduced by storing data in the cloud rather than transferring it from outside.*

## 1. Introduction

The developers of scientific applications have many options when it comes to choosing a platform to run their applications. In the past these options included: local workstations, clusters, supercomputers and grids. Each of these choices offers various tradeoffs in terms of usability, performance, and cost. Recently, *cloud computing* has emerged as another promising solution for scientific applications and is rapidly gaining interest in the scientific community.

Many definitions of cloud computing have been proposed [4,34] [13]. These definitions vary in the scope of what constitutes a cloud and what features a cloud provides. For the purposes of this paper we consider a cloud to be a cluster that offers virtualized computational resources, service-oriented provisioning, and a "pay as you go" usage-based pricing model. Currently there are several commercial clouds that offer these features, such as Amazon EC2 [2], GoGrid [17], and FlexiScale [12]. In addition, it is now possible to build private clouds using open-source cloud computing middleware such as Eucalyptus [27], OpenNebula [28], and Nimbus [26].

Clouds offer many technical and economic advantages over other platforms that are just beginning to be identified. They combine the customization of virtual machines, the scalability and resource sharing of grids, and the stability and economy of software as a service (SaaS). The use of virtualization in particular has been shown to provide many useful benefits for scientific applications, including: user-customization of system software and services, performance isolation, check-pointing and migration, better reproducibility of scientific analyses, and enhanced support for legacy applications [11][19].

Recently, many studies have investigated the use of clouds and virtualization for scientific applications [33] [36] [24] [32] [35] [10] [16]. These studies have primarily focused on tightly-coupled applications and common HPC benchmarks.

In this paper we study the use of cloud computing for scientific workflows. Workflows are loosely-coupled parallel applications that consist of a series of computational tasks connected by data- and control-flow dependencies. Many scientific analyses are easily expressed as workflows and they are commonly used to solve problems in many disciplines [37].

Clouds provide several benefits for workflow applications. These benefits include:

- **Illusion of infinite resources**—Unlike grids, clouds give the illusion that the available computing resources are unlimited. This means that users can request, and are likely to obtain, sufficient resources at any given time. Existing commercial clouds have a different workload than grids, however, and the illusion may break down for very large workflows, or if clouds become popular for scientific computing.
- **Leases**—In grids and clusters the user specifies the amount of time required for a computation and delegates responsibility for allocating resources to a batch scheduler. In clouds, the user directly allocates resources as required to schedule their computations. This model is ideal for workflows and other loosely-coupled applications because it decreases the scheduling overheads that can significantly reduce their performance.
- **Elasticity**—Clouds allow users to acquire and release resources on-demand. This enables workflow systems to easily grow and shrink the available resource pool as the needs of the workflow change over time.

Previous work on the use of cloud computing for workflows has studied the cost and performance of clouds via simulation [8] and using an experimental cloud [18]. In this paper we extend that work using several workflows that represent different domains and different resource requirements. We use an existing commercial cloud, Amazon's EC2 [2], in order to assess the potential of currently deployed clouds. We analyze the cost of running the experiments on EC2, and compare the EC2 performance to a typical HPC system, NCSA's Abe cluster [25].

The contributions of this paper are:
- an experimental study of the performance of three workflows with different I/O, memory and CPU requirements on a commercial cloud
- a comparison of the performance of cloud resources and typical HPC resources, and
- an analysis of the various costs associated with running workflows on a commercial cloud.

In this paper we focus on single, multi-core node performance, which provides adequate capabilities for the applications we are evaluating in this paper.

## 2. Applications

In order to evaluate the usefulness of cloud computing for scientific workflows we ran three different workflow applications: an astronomy application (Montage), a seismology application (Broadband), and a bioinformatics application (Epigenomics). These three applications were chosen because they cover a wide range of application domains and a wide range of resource requirements. Table 1 shows the relative resource usage of these applications in three different categories: I/O, memory, and CPU. In general, applications with high I/O usage are I/O-bound, applications with high memory usage are memory-limited, and applications with high CPU usage are CPU-bound.

Montage [21] creates science-grade astronomical image mosaics using data collected from telescopes. The size of a Montage workflow depends upon the area of the sky (in square degrees) covered by the output mosaic. In our experiments we configured Montage workflows to generate an 8-degree mosaic. The resulting workflow contains 10,429 tasks, reads 4.2 GB of input data, and produces 7.9 GB of output data. Montage is considered to be I/O-bound because it spends more than 95% of its time waiting on I/O operations.

Broadband [30] generates and compares seismograms from several high- and low-frequency earthquake simulation codes. Each workflow generates seismograms for several sources (scenario earthquakes) and sites (geographic locations). For each (source, site) combination the workflow runs several high- and low-frequency earthquake simulations and computes intensity measures of the resulting seismograms. In our experiments we used 4 sources and 5 sites to generate a workflow containing 320 tasks that reads 6 GB of input data and writes 160 MB of output data. Broadband is considered to be memory-limited because more than 75% of its runtime is consumed by tasks requiring more than 1 GB of physical memory.

Epigenome [31] maps short DNA segments collected using high-throughput gene sequencing machines to a previously constructed reference genome using the MAQ software [22]. The workflow splits several input segment files into small chunks, reformats and converts the chunks, maps the chunks to a reference genome, merges the mapped sequences into a single output map, and computes the sequence density for each location of interest in the reference genome. The workflow used in our experiments maps human DNA sequences to a reference chromosome 21. The workflow contains 81 tasks, reads 1.8 GB of input data, and produces 300 MB of output data. Epigenomics is considered to be CPU-bound because it spends 99% of its runtime in the CPU and only 1% on I/O and other activities.

**Table 1**: Application resource usage comparison

| Application | I/O | Memory | CPU |
|---|---|---|---|
| Montage | High | Low | Low |
| Broadband | Medium | High | Medium |
| Epigenomics | Low | Medium | High |

## 3. Execution Environment

In this section we describe the experimental setup that was used to run workflows. We ran experiments on Amazon EC2 and NCSA's Abe cluster. EC2 was chosen because it is currently the most popular, feature-rich, and stable commercial cloud. Abe was chosen because it is typical of the existing HPC systems a scientist could choose to run their workflow application and is therefore a logical alternative to EC2.

Workflows are loosely-coupled parallel applications that consist of a set of computational tasks linked via data- and control-flow dependencies. Unlike tightly-coupled applications in which tasks communicate directly via the network, workflow tasks typically communicate using the file system. Each task produces one or more output files that become input files to other tasks. In HPC systems these files are typically stored on a network file system, which allows the workflow to run in parallel on several nodes.

**Table 2:** Resource types used

| Type | Arch. | CPU | Cores | Memory | Network | Storage | Price |
|---|---|---|---|---|---|---|---|
| m1.small | 32-bit | 2.0-2.6 GHz Opteron | 1/2 | 1.7 GB | 1-Gbps Ethernet | Local disk | $0.10/hr |
| m1.large | 64-bit | 2.0-2.6 GHz Opteron | 2 | 7.5 GB | 1-Gbps Ethernet | Local disk | $0.40/hr |
| m1.xlarge | 64-bit | 2.0-2.6 GHz Opteron | 4 | 15 GB | 1-Gbps Ethernet | Local disk | $0.80/hr |
| c1.medium | 32-bit | 2.33-2.66 GHz Xeon | 2 | 1.7 GB | 1-Gbps Ethernet | Local disk | $0.20/hr |
| c1.xlarge | 64-bit | 2.33-2.66 GHz Xeon | 8 | 7.5 GB | 1-Gbps Ethernet | Local disk | $0.80/hr |
| abe.local | 64-bit | 2.33 GHz Xeon | 8 | 8 GB | 10-Gbps InfiniBand | Local disk | N/A |
| abe.lustre | 64-bit | 2.33 GHz Xeon | 8 | 8 GB | 10-Gbps InfiniBand | Lustre | N/A |

One of the advantages of HPC systems over currently deployed commercial clouds is the availability of high-performance I/O devices. HPC systems commonly provide high-speed networks and parallel file systems, while most commercial clouds use commodity networking and storage devices. These high-performance devices increase workflow performance by making inter-task communication more efficient. In order to have an unbiased comparison of the performance of workflows on EC2 and Abe, the experiments presented in this paper attempt to account for these differences by a) running all experiments on single nodes and b) running experiments using the local disk on both EC2 and Abe, and the parallel file system on Abe.

The use of single nodes minimizes the advantage that Abe has as a result of having a high-speed interconnect. Although single-node experiments do not enable us to measure the scalability of cloud services they do provide an application-oriented understanding of the capabilities of the underlying resources that can help in making provisioning decisions. Testing the scalability of cloud services when running workflows on multiple nodes is left for future work.

Running experiments using both the parallel file system and the local disk on Abe allows us to determine what performance advantage, if any, Abe nodes have as a result of parallel I/O. It is expected that the use of a parallel file system will significantly improve the runtime of I/O-intensive applications like Montage, but will be less of an advantage for CPU-intensive applications like Epigenome.

### 3.1 Resources

Table 2 compares the resource types used for the experiments. It lists 5 resource types from EC2 (m1.* and c1.*) and 2 resource types from Abe (abe.local and abe.lustre). There are several noteworthy details about the resources shown. First, although there is actually only one type of Abe node, there are two types listed in the table: abe.local and abe.lustre. The actual hardware used for these types is equivalent, the difference is in how I/O is handled. The abe.local type uses a local partition for I/O, and the abe.lustre type uses a Lustre partition for I/O. Using two different names is simply a notational convenience.

Second, in terms of computational capacity, the c1.xlarge resource type is roughly equivalent to the abe.local resource type with the exception that abe.local has slightly more memory. We use this fact to estimate the virtualization overhead for our test applications on EC2.

Third, in rare cases EC2 assigns Xeon processors for m1.* instances, but for all of the experiments reported here the m1.* instances used were equipped with Opteron processors. The only significance is that Xeon processors have better floating-point performance than Opteron processors (4 FLOP/cycle vs. 2 FLOP/cycle).

Finally, the m1.small instance type is shown having ½ core. This is possible because of virtualization. EC2 nodes are configured to give m1.small instances access to the processor only 50% of the time. This allows a single processor core to be shared equally between two separate m1.small instances.

### 3.2 Software

All workflows were planned and executed using the Pegasus Workflow Management System [9] with DAGMan [6] and Condor [23]. Pegasus is used to transform abstract workflow descriptions into concrete plans, which are then executed using DAGMan to manage task dependencies, and Condor to manage task execution.

The software was deployed on EC2 as shown in Figure 1. A submit host running outside the cloud was used to coordinate the workflow, and worker nodes were started inside the cloud to execute workflow tasks. Two virtual machine images were used to start worker nodes: one for 32-bit instance types and one for 64-bit instance types. Both images were based on the standard Fedora Core 8 images provided by Amazon. To the base images we added Condor, Pegasus and other miscellaneous packages required to compile and run the selected applications. Compressed, the 32-bit image was 773 MB and the 64-bit image was 729 MB. Uncompressed, the 32-bit image was 2.1 GB and the 64-bit image was 2.2 GB. The images did not include any application-specific configurations, so we were able to use the same set of images for all experiments. All images are stored in Amazon S3 [3]. S3 is an object-based, replicated storage service that supports simple PUT and GET operations on file-like binary objects.

For the Abe experiments Globus [14] and Corral [7] were used to deploy Condor glideins [15] as shown in Figure 2. The glideins started Condor daemons on the Abe worker nodes, which contacted the submit host and were used to execute workflow tasks. This approach creates an execution environment on Abe that is equivalent to the EC2 environment.

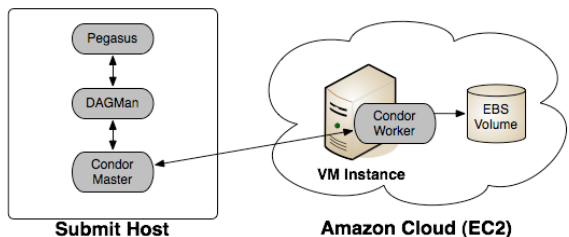

**Figure 1**: Execution environment on EC2

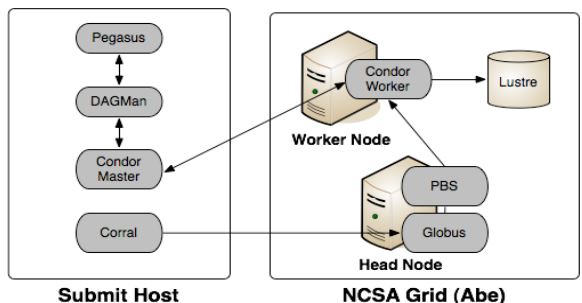

**Figure 2**: Execution environment on Abe

### 3.3 Storage

To run workflows we need to allocate storage for 1) application executables, 2) input data, and 3) intermediate and output data. In a typical workflow application executables are pre-installed on the execution site, input data is copied from an archive to the execution site, and output data is copied from the execution site to an archive. For these experiments, executables and input data were pre-staged to the execution site, and output data were not transferred from the execution site.

For EC2, executables were installed in the VM images, intermediate and output data was written to a local partition, and input data was stored on EBS volumes.

The Elastic Block Store (EBS) [1] is a SAN-like, replicated, block-based storage service that can be used with EC2 instances. EBS volumes can be created in any size between 1 GB and 1 TB and appear as standard, unformatted block devices when attached to an EC2 instance. As such, EBS volumes can be formatted with standard UNIX file systems and used like an ordinary disk, but they cannot be shared between multiple instances.

EBS was chosen to store input data for a number of reasons. First, storing inputs in the cloud obviates the need to transfer input data repeatedly. This saves both time and money because transfers cost more than storage. Second, using EBS avoids the 10 GB limit on VM images, which is too small to include the input data for all the applications tested. We can access the data as if it were on a local disk without packaging it in the VM image. A simple experiment using the disk copy utility 'dd' showed similar performance reading from EBS volumes and the local disk (74.6 MB/s for local, and 74.2 MB/s for EBS). Finally, using EBS simplifies our setup by allowing us to reuse the same volume for multiple experiments. When we need to change instances we just detach the volume from one instance and re-attach it to another.

For Abe, all application executables and input files were stored in the Lustre file system. For abe.local experiments the input data was copied to a local partition (/tmp) before running the workflow, and all intermediate and output data was written to the same local partition. For abe.lustre, all intermediate and output data was written to the Lustre file system.

### 4. Performance Comparison

In this section we compare the performance of the selected workflow applications by executing them on Abe and EC2. The critical performance metric we are concerned with is the runtime of the workflow (also known as the makespan), which is the total amount of wall clock time from the moment the first workflow task is submitted until the last task completes. The runtimes reported for EC2 do not include the time required to install and boot the VM, which typically averages between 70 and 90 seconds [20], and the runtimes reported for Abe do not include the time glidein jobs spend waiting in the queue, which is highly dependent on the current system load. Also, the runtimes do not include the time required to transfer input and output data (see Table 4). We assume that this time will be variable depending on WAN conditions. A study of bandwidth to/from Amazon services is presented in [29]. In our experiments we typically observed bandwidth on the order of 500-1000KB/s between EC2 and our submit host in Marina del Rey, CA.

We estimate the virtualization overhead for each application by comparing the runtime on c1.xlarge with the runtime on abe.local. Measuring the difference in runtime between these two resource types should provide a good estimate of the cost of virtualization.

Figure 3 shows the runtime of the selected applications using the resource types shown in Table 2. In all cases the m1.small resource type had the worst runtime by a large margin. This is not surprising given its relatively low capabilities.

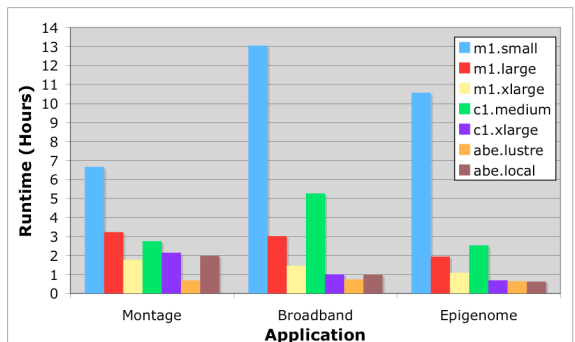

**Figure 3:** Runtime comparison

### 4.1 Montage

For Montage the best EC2 performance was achieved on the m1.xlarge type. This is likely due to the fact that m1.xlarge has twice as much memory as the next best resource type. The extra memory is used by the Linux kernel for the file system buffer cache to reduce the amount of time the application spends waiting for I/O. This is particularly beneficial for Montage, which is very I/O-intensive.

The best overall performance for Montage was achieved using the abe.lustre configuration, which was more than twice as fast as abe.local. This large gap suggests that having a parallel file system is a significant advantage for I/O-bound applications like Montage.

The difference in runtime between the c1.xlarge and abe.local experiments suggests that the virtualization overhead for Montage is less than 8%.

### 4.2 Broadband

The best overall runtime for Broadband was achieved by using the abe.lustre resource type, and the best EC2 runtime was achieved using the c1.xlarge resource type. This is despite the fact that only 6 of the 8 cores on c1.xlarge and abe.lustre could be used due to memory limitations.

Unlike Montage the difference between running Broadband on a relatively slow local disk (abe.local) and running on the parallel file system (abe.lustre) is not as significant. This is attributed to the lower I/O requirements of Broadband.

Broadband performs the worst on m1.small and c1.medium, which also have the lowest amount memory (1.7 GB). This is because m1.small has only half a core, and c1.medium can only use one of its two cores because of memory limitations.

The difference between the runtime using c1.xlarge and the runtime using abe.local was only about 1%. This small difference suggests a relatively low virtualization penalty for Broadband.

### 4.3 Epigenomics

For Epigenomics the best EC2 runtime was achieved using c1.xlarge and the best overall runtime was achieved using abe.lustre. The primary factor affecting the performance of Epigenome was the availability of processor cores, with more cores resulting in a lower runtime. This is expected given that Epigenome is almost entirely CPU-bound.

The difference between the abe.lustre and abe.local runtimes was only about 2%, which is consistent with the fact that Epigenome has relatively low I/O and is therefore less affected by the parallel file system.

The difference between the abe.local and the c1.xlarge runtimes suggest that the virtualization overhead for this application is around 10%, which is higher than both Montage and Broadband. This may suggest that virtualization has a larger impact on CPU-bound applications.

## 5. Cost Analysis

In this section we analyze the cost of running workflow applications in the cloud. We consider three different cost categories: resource cost, storage cost, and transfer cost. Resource cost includes charges for the use of VM instances in EC2; storage cost includes charges for keeping VM images in S3 and input data in EBS; and transfer cost includes charges for moving input data, output data and log files between the submit host and EC2.

### 5.1 Resource Cost

Each of the five resource types Amazon offers is charged at a different hourly rate: $0.10/hr for m1.small, $0.40/hr for m1.large, $0.80/hr for m1.xlarge, $0.20/hr for c1.medium, and $0.80/hr for c1.xlarge. Usage is rounded up to the nearest hour, so any partial hours are charged as full hours.

Figure 4 shows the per-workflow resource cost for the applications tested. Although it did not perform the best in any of our experiments, the most cost-effective instance type was c1.medium, which had the lowest execution cost for all three applications.

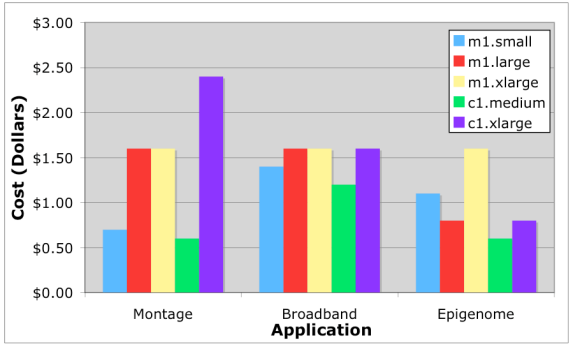

**Figure 4:** Resource cost comparison

*5.2 Storage Cost*

Storage cost consists of a) the cost to store VM images in S3, and b) the cost of storing input data in EBS. Both S3 and EBS use fixed monthly charges for the storage of data, and variable usage charges for accessing the data. The fixed charges are $0.15 per GB-month for S3, and $0.10 per GB-month for EBS. The variable charges are $0.01 per 1,000 PUT operations and $0.01 per 10,000 GET operations for S3, and $0.10 per million I/O operations for EBS. We report the fixed cost per month, and the total variable cost for all experiments performed.

We used a 32-bit and a 64-bit VM image for all of the experiments in this paper. The 32-bit image was 773 MB and the 64-bit image was 729 MB for a total fixed cost of $0.22 per month. In addition, there were 4616 GET operations and 2560 PUT operations for a total variable cost of approximately $0.03.

The fixed monthly cost of storing input data for the three applications is shown in Table 3. In addition, there were 3.18 million I/O operations for a total variable cost of $0.30.

**Table 3:** Monthly storage cost

| Application | Volume Size | Monthly Cost |
|---|---|---|
| Montage | 5GB | $0.66 |
| Broadband | 5GB | $0.60 |
| Epigenome | 2GB | $0.26 |

*5.3 Transfer Cost*

In addition to resource and storage charges, Amazon charges $0.10 per GB for transfer into, and $0.17 per GB for transfer out of, the EC2 cloud. Tables 4 and 5 show the per-workflow transfer sizes and costs for the three applications studied. Input is the amount of input data to the workflow, output is the amount of output data, and logs is the amount of logging data that is recorded for workflow tasks and transferred back to the submit host. The cost of the protocol used by Condor to communicate between the submit host and the workers is not included, but it is estimated to be less than $0.01 per workflow.

**Table 4:** Per-workflow transfer sizes

| Application | Input | Output | Logs |
|---|---|---|---|
| Montage | 4291 MB | 7970 MB | 40 MB |
| Broadband | 4109 MB | 159 MB | 5.5 MB |
| Epigenome | 1843 MB | 299 MB | 3.3 MB |

**Table 5:** Per-workflow transfer costs

| Application | Input | Output | Logs | Total |
|---|---|---|---|---|
| Montage | $0.42 | $1.32 | < $0.01 | $1.75 |
| Broadband | $0.40 | $0.03 | < $0.01 | $0.43 |
| Epigenome | $0.18 | $0.05 | < $0.01 | $0.23 |

## 6. Discussion

*6.1 Performance*

Based on these experiments we believe the performance of workflows on EC2 is reasonable given the resources that can be provisioned. Although the EC2 performance was not as good as the performance on Abe, most of the resources provided by EC2 are also less powerful. In the cases where the resources are similar, the performance was found to comparable. The EC2 c1.xlarge type, which is nearly equivalent to abe.local, delivered performance that was nearly the same as abe.local in our experiments.

For I/O-intensive workflows like Montage, EC2 is at a significant disadvantage because of the lack of high-performance parallel file systems. While such a file system could conceivably be constructed from the raw components available in EC2, the cost of deploying such a system would be prohibitive. In addition, because EC2 uses commodity networking equipment it is unlikely that there would be a significant advantage in shifting I/O from a local partition to a parallel file system across the network, because the bottleneck would simply shift from the disk to the network interface. In order to compete performance-wise with Abe for I/O-intensive applications, Amazon would need to deploy both a parallel file system and a high-speed interconnect.

For memory-intensive applications like Broadband, EC2 can achieve nearly the same performance as Abe as long as there is more than 1 GB of memory per core. If there is less, then some cores must sit idle to prevent the system from running out of memory or swapping. This is not strictly an EC2 problem, the same issue affects Abe as well.

For CPU-intensive applications like Epigenome, EC2 can deliver comparable performance given equivalent resources. The virtualization overhead does not seem to be a significant barrier to performance for such applications. In fact, the virtualization overhead measured for all application less than 10%. This is consistent with previous studies that show similar virtualization overheads [5,16,36]. As such, virtualization does not seem, by itself, to be a significant performance problem for clouds. As virtualization technologies improve it is likely that what little overhead there is will be further reduced or eliminated.

*6.2 Cost*

The first thing to consider when provisioning resources on EC2 is the tradeoff between performance and cost. In general, EC2 resources obey the aphorism "you get what you pay for"—resources that cost more perform better than resources that cost less. For the applications tested, c1.medium was the most cost-effective resource type even though it did not have the lowest hourly rate, because the type with the lowest rate (m1.small) performed so badly.

Another important thing to consider when using EC2 is the tradeoff between storage cost and transfer cost. Users have the option of either a) transferring input data for each workflow separately, or b) transferring input data once, storing it in the cloud, and using the stored data for multiple workflow runs. The choice of which approach to employ will depend on how many times the data will be used, how long the data will be stored, and how frequently the data will be accessed. In general, storage is more cost-effective for input data that is reused often and accessed frequently, and transfer is more cost-effective if data will be used only once. For the applications tested in this paper, the monthly cost to store input data is only slightly more than the cost to transfer it once. Therefore, for these applications, it is usually more cost-effective to store the input data rather than transfer the data for each workflow.

Although the cost of transferring input data can be easily amortized by storing it in the cloud, the cost of transferring output data may be more difficult to reduce. For many applications the output data is much smaller than the input data, so the cost of transferring it out may not be significant. This is the case for Broadband and Epigenome, for example. For other applications the large size of output data may be cost-prohibitive. In Montage, for example, the output is actually larger than the input and costs nearly as much to transfer as it does to compute. For these applications it may be possible to leave the output in the cloud and perform additional analyses there rather than to transfer it back to the submit host.

In [8] the cost of running 1-, 2-, and 4-degree Montage workflows on EC2 was studied via simulation. That paper found the lowest total cost of a 1-degree workflow to be $0.60, a 2-degree to be $2.25, and a 4-degree to be $9.00. In comparison, we found the total cost of an 8-degree workflow, which is 4 times larger than a 4-degree workflow, to be approximately $1.25 if data is stored for an entire month, and $2.35 if data is transferred. This difference is primarily due to an underestimate of the performance of EC2 that was used in the simulation, which produced much longer simulated runtimes.

Finally, the total cost of all the experiments presented in this paper was $149.55. That includes all charges related to learning to use EC2, creating VM images, and running test and experimental workflows.

## 7. Conclusion

In this paper we examined the performance and cost of running scientific workflow applications in the cloud using Amazon's EC2 as a model. We ran several workflow applications representing diverse application domains and resource requirements on EC2 and compared the performance to NCSA's Abe cluster. We found that although the performance of EC2 was not equivalent to Abe in most cases, it was reasonable given the resources available. The primary advantages of Abe were found to be the availability of a high-speed interconnect, and a parallel file system, which significantly improved the performance of the I/O-intensive application. Factoring out these advantages by running additional Abe tests using the local disk shows that, given equivalent resources, EC2 is capable of performance close to that of Abe. All other things being equal the only difference was a small virtualization overhead in EC2, which was measured to be between 1% and 10% for the applications tested.

We also analyzed the cost of running workflows on EC2. We found that the primary cost was in acquiring resources to execute workflow tasks, and that storage costs were relatively small in comparison. The cost of data transfers, although relatively high, can be effectively reduced by storing data in the cloud rather than transferring it for each workflow. In addition, we found the cost of running workflows in the cloud to be much less, and the performance to be much better, than suggested by previous research.

These results indicate that clouds are a viable alternative for running scientific workflow applications, but unless cloud providers begin offering high-speed networks and parallel file systems they are unlikely to compete with existing HPC systems in terms of performance.

In this paper we focused on the case where only a single node is used to run a workflow. In the future we plan to extend this work to study the performance and cost of clouds when multiple nodes are used. That study will include an analysis of the various ways in which data can be communicated between workflow tasks in a cloud.

## Acknowledgements

This work was supported by the National Science Foundation under the SciFlow (CCF-0725332) grant. This research made use of Montage, funded by the National Aeronautics and Space Administration's Earth Science Technology Office, Computation Technologies Project, under Cooperative Agreement Number NCC5-626 between NASA and the California Institute of Technology.